\documentclass[12pt]{article}
\newcommand{\be}{\begin{eqnarray}}
\newcommand{\ee}{\end{eqnarray}}

\newcommand{\ovl}{\overline}

\begin{document}

\begin{titlepage}

\begin{flushright}
hep-ph/0403082
\end{flushright}
\vskip 2.5cm

\begin{center}
{\Large \bf
Decay properties of new $D$-mesons}
\end{center}

\vspace{1ex}

\begin{center}
{\large Ya.I. Azimov$^{a,b}$, K. Goeke$^b$}

\vspace{5mm}
{\sl ${}^a$~Petersburg Nuclear Physics Institute,\\
St.Petersburg, 188300 Russia~~~~~~~~~ \\
 ${}^b$~Institute for Theoretical Physics II,~~~ \\
   ~~~~~~~~~~~Ruhr University, 44801 Bochum, Germany
} \\

\end{center}

\vspace{2.5ex}

\medskip

\centerline {\bf Abstract}

\bigskip

We consider radiative and pionic decays of the new $D_s$-mesons
in the framework of a phenomenologically motivated approach.
Present data on ratios of the two kinds of decays can be described
without explicit using a 4-quark component. Most probably, the
isospin violation in decays of different $D_s$-mesons is not
universal, and the binding potential should be different from
Coulombic. New precise measurements may provide further
clarification for the nature of the $D_s$ excited states.

\bigskip

\end{titlepage}

\section{Introduction}

Recent discoveries of several new charmed meson states are among
the most impressive recent events in the heavy flavour physics.
These are, first of all, two narrow charmed-strange states,
$D_{sJ}(2317)$ and $D_{sJ}(2460)\,$. The former one was initially
observed by the BaBar group~\cite{BaB1}, and later confirmed by
CLEO~\cite{CLEO} and Belle~\cite{Bel1,Bel2}. Some evidence for the
second state was also noted in ref.~\cite{BaB1}, but it was seen
definitely only later, by CLEO~\cite{CLEO} and Belle~\cite{Bel1,Bel2}.
Further, it was reliably confirmed by BaBar as well~\cite{BaB2}.
To the present moment the two states have been seen both in
$B$-decays and in the continuum annihilation $e^+e^-\to c\,
\ovl c $~\cite{Bel1}.  So they are definitely established,
though their properties, especially their quantum numbers, are not
quite clear yet (experimental minireviews on these states see
in~\cite{FPort,Wang}).

Two more states, non-strange charmed mesons, seem to be discovered
by Belle~\cite{Bel3} in final states $D^{\pm}\pi^{\mp}$ and
$D^{*\pm}\pi^{\mp}\,$. They have been extracted from the coherent
amplitude analysis of the Dalitz plots for $D\pi\pi$ and $D^*\pi\pi\,$.
As a result, their quantum numbers are fixed (dominant $S$-waves),
but their existence strongly needs confirmation. It is even more so,
that these states have very large widths (some hundreds MeV), and
overlap with kinematical reflections from other resonances. Evidence
for the broad $S$-wave state in the $D\pi$-system has been recently
presented also by the Wideband photoproduction experiment
FOCUS~\cite{foc}. However, the masses as given by the two
collaborations differ by a couple of standard deviations,
and the situation needs further clarification.

Unexpected properties of the new $D_s$-states have induced active
theoretical discussions. Presumably, the mesons have spin-parity
numbers $0^+$ for $D_{sJ}(2317)$ and $1^+$ for $D_{sJ}(2460)$
(these values, at least, do not contradict any experimental data).
States with such quantum numbers have been, of course, predicted
earlier, but they were expected to have higher masses and much
higher widths~\cite{BS,DPE}. Therefore, various suggestions have
appeared in the literature that the new states may have some
nature different from the canonical  $q\ovl q$ states.
The published ideas include $DK$-molecules~\cite{DKmol},
$D\pi$-atom~\cite{szcz}, 4-quark states~\cite{4q}, and their
combination with $q\ovl q$ states~\cite{TBr}.

In the framework of more ``usual'' approaches, the chiral
perturbation theory was combined with HQET to include the new
states into $D_s$ spectroscopy as chiral partners of the
well-known states $D_s(0^-)$ and
$D_s^*(1^-)$~\cite{WB,Now1,Deand}. The previous calculations for
masses were mainly based on the relativistic quark model with the
potential consisting of two parts: Coulomb interaction at small
distances (due to one-gluon exchange), and linear potential at large
distances (for confinement)~\cite{GI,GK,EHQ}. Now it was suggested
that such method could be applied to masses of the new states as
well, if one modified the confinement potential~\cite{CJ}.

Decay properties of the new states have been also considered in
various assumptions, {\it e.g.}, HQET with chiral effective
Lagrangian for light quarks~\cite{WB,CDF}, previous~\cite{God} and
modified~\cite{CJ} potential pictures. However, all they show
disagreement with present experimental data, especially for the
ratio of radiative and hadronic decays of the new mesons. This has
led to the rather pessimistic conclusion that if the data are
confirmed ``the conventional $D_{s0}^*$ and $D_{s1}^{1/2}$ states
have yet to be discovered''~\cite{God}.

In this note we use another, essentially phenomenological way to
estimate partial widths of the new $D_s$ states. We try to use as
few dynamical assumptions as possible. Our approach is explained
in Section 2. Then it is applied to the new meson decays (Sections
3,4). Our results are summarized and discussed in Section 5.

In what follows we always assume that the states $D_{sJ}(2317)$
and $D_{sJ}(2460)$ have indeed the spin-parity $0^+$ and $1^+$
respectively.  We will denote them as $D_{s0}^*(2317)$ and
$D_{s1}(2460)\,$.

\section{Mass dependence of matrix elements}

To explain our approach, let us discuss the problem in terms of
2-body configurations with an efficient potential. Then decay
amplitudes are expressible through some overlap integrals between
initial and final wave functions.

For the pure Coulomb potential, having no specific mass-dimensional
parameter, the overlap integrals could be written as a power of
the reduced mass $\mu=m_1\,m_2/(m_1+ m_2)$ multiplied by a universal
dimensionless factor (independent of the constituent masses). The
same would be true, in particular, for mass differences of radial
excitations having the same flavour quantum numbers. Those mass
differences would be proportional to the corresponding reduced masses.

Experimental data, however, do not support such expectations.
Let us take, for instance, vector and tensor mesons with different
flavour content. They are well established and well studied for all
known flavours (excluding beauty mesons), they nicely correspond to
expectations for the quark-antiquark system. The both states
have the same internal spin structure (the total spin of constituents
is $S=1$), and differ only in the internal orbital momentum
($L=0,\,1$ for the vector and tensor states respectively). Moreover,
tensor and vector non-strange isospin-singlet mesons have nearly
the same $SU(3)_F$ violating mixing angles, leading to nearly ideal
separation of the strange and non-strange components. It is
reasonable, therefore, to compare tensor-vector mass differences for
mesons of various flavours.

For the mesons with light ($u,d,s$) quarks we have the
differences~\cite{PDG}
\be \label{Mtv-uds}
M_{a_2}-M_{\rho}=(547\pm1)~{\rm MeV}\,,\\
M_{K_2^*}-M_{K^*}=(534\pm2)~{\rm MeV}\,,\\
M_{f_2}-M_{\omega}=(493\pm1)~{\rm MeV}\,,\\
M_{f'_2}-M_{\phi}=(505\pm5)~{\rm MeV}\,, \ee
which are nearly the same. Note that the change $(u,d)\to s\,$,
though essentially shifts the particle masses, produces only
very small effect ($\sim 2.5\%$) on $T-V$ mass differences.

For mesons with heavy quarks the differences are~\cite{PDG}
\be \label{Mtv-cb}
M_{D_2^{*0}}-M_{D^{*0}}=(452\pm2)~{\rm MeV}\,,\\
M_{D_2^{*\pm}}-M_{D^{*\pm}}=(449\pm4)~{\rm MeV}\,,\\
M_{D_{s2}^{*\pm}}-M_{D_s^{*\pm}}=(460\pm2)~{\rm MeV}\,,\\
M_{\chi_{c2}}-M_{J/\psi}=(459.31\pm0.14)~{\rm MeV}\,,\\
M_{\chi_{b2}(1P)}-M_{\Upsilon(1S)}=(452.3\pm0.6)~{\rm MeV}\,. \ee
Again, the change of the mass difference under substitution
$(u,d)\to s$ is only $\sim 2.5\%\,$. With this accuracy the $T-V$
mass differences are the same for all mesons containing at least
one heavy quark, $c$ or $b\,$, being lower(!) than ones for light-quark
mesons\footnote{Recall that $T$- and $V$-mesons differ only in the
internal orbital momentum. Mass difference for such systems bound by
the Coulomb potential should grow with increase of constituent
mass(es). In contrast, $V$- and $P$-mesons differ only in spin
orientation, and their mass differences should decrease with
increasing constituent masses, independently of the binding potential,
due to decrease of (chromo)magnetic moments. This latter expectation
exactly agrees with measurements.}. It is interesting to note
that in the sequence of hidden-flavour quarkonia $[f_2,\omega]\to
[f'_2,\phi]\to[\chi_{c2},J/\psi]\to[\chi_{b2}(1P),\Upsilon(1S)]\,$,
which corresponds to the quark sequence $u/d\to s\to c\to b$ with
monotonously growing constituent masses, the $T-V$ mass difference
slightly increases at the first step, and then monotonously decreases.
Such behaviour does not correspond to Coulomb regularities, even for
the most compact, $b$-quark systems.

The discussed $T-V$ mass differences are free from the constituent
``rest mass'' contributions, and are expressed only through diagonal
matrix elements of some operators (corresponding to the kinetic and
potential energy). As we see, such matrix elements, for changing quark
contents, vary not more than $\sim20\%\,$, {\it i.e.}, they reveal only
low dependence on the quark masses.

Decay amplitudes are also expressed through some matrix elements,
though nondiagonal ones, and of different operators. Nevertheless,
we will assume that the emerging matrix elements also have low
dependence on the quark masses. As the first approximation, we will
neglect such dependence.

\section{Decays of $D_{s0}(2317)$}

Let us begin with consideration of $D_{s0}^*(2317)\,$. Its only
observed decay mode is \be \label{s0pi} D_{s0}^{*\pm}(2317)\to
D_s^{\pm}\,\pi^0\,, \ee
having the final momentum about 320~MeV/c. If the initial meson
corresponds to the $c\ovl s$ system, then this decay violates
isospin, and should have some suppression. It is well known that
there are at least two possible sources for such violation,
electromagnetic interactions and/or mass difference of $d$- and
$u$-quarks. Examples of the isospin violation for light-quark
hadrons (without explicit photon radiation or absorption)
demonstrate numerical smallness of order $10^{-2}-10^{-3}
\sim\cal O(\alpha)$ (the rare exclusion, rather intensive decay
$\omega\to\pi^+\pi^-\,$, only confirms this empirical rule, since
contains enhancement due to the small $\rho\omega$ mass difference).
This does not allow to separate strong and electromagnetic
violations of the isospin.

Situation for heavy-quark hadrons seems to be different. The best
studied example is the isospin-forbidden decay $\psi(2S)\to
J/\psi\,\pi^0$ with branching ratio~\cite{PDG}
$$B_{\pi^0}=(9.6\pm2.1)\cdot10^{-4}$$ and final momentum
$p_{\pi^0}=527$~MeV/c. It can be compared with the similar, but
isospin-allowed decay $\psi(2S)\to J/\psi\,\eta$ having branching
ratio~\cite{PDG} $$B_{\eta}=(3.17\pm0.21)\cdot10^{-2}$$ and the
final momentum $p_{\eta}=200$~MeV/c. Both decays produce $p$-wave
final states. So, the suppression factor in the isospin-violating
amplitude is
\be \epsilon=\left[\frac{B_{\pi^0}}{B_{\eta}}
\left(\frac{p_{\eta}}{p_{\pi^0}}\right)^3\right]^{1/2}
=(4.07\pm0.47)\cdot10^{-2}\,. \label{supr}\ee
Such a value looks too large for the electromagnetic mechanism. 
It might be enhanced by $\eta\,\pi^0$-mixing, in similarity 
(though weaker) with $\rho^0\,\omega$-mixing which enhances the 
isospin-violating decay $\omega\to\pi^+\pi^-$.  If so, the same 
parameter $\epsilon$ should work in all decays which produce $\pi^0$ 
with violation of isospin. Alternatively, its large value in compact 
charmonium mesons could give evidence for the direct effect of the mass 
difference $m_d-m_u$, being determined, say, by the ratio 
$(m_d-m_u)/\Lambda_{\rm QCD}\,$.  In any case, we will try to assume 
that this parameter is universally applicable to isospin violation in 
decays of both charmonium and $D_s$-mesons.

Now we can compare decay (\ref{s0pi}) with the similar, but
isospin-conserving decay
\be \label{d0pi} D_{0}^{*\pm}\to D^{\pm}\,\pi^0 \,,\ee
where $D_0^{*\pm}$ are charmed-nonstrange mesons with
$J^P=0^+\,.$ Their isotopic partner, $D_0^{*0}\,$, was found
by Belle~\cite{Bel3} to have the mass and total width
\be M_0^*=(2308\pm37)~{\rm MeV}\,,~~~~\Gamma_0^*= (276\pm66)~{\rm
MeV}\,. \label{d0bel} \ee
With reasonable accuracy, we can assume for $D_{0}^{*\pm}$
the same values of the mass and total width. Further, we suggest
that the large total width is completely due to the decay mode
$D_{0}^{*}\to D\pi$ which should be strongly dominant. For the
particular charged channel (\ref{d0pi}) we then obtain the final
momentum about 378~MeV/c and the partial width
\be \Gamma(D_{0}^{*\pm}\to D^{\pm}\,\pi^0)\approx(1/3)
\Gamma_0^* =(92\pm22)~{\rm MeV}\,.\ee
The rest $(2/3)\Gamma_0^*$ comes, according to isotopic relations,
from the mode $D_{0}^{*\pm}\to D^{0}\,\pi^{\pm}\,$.

The both decays (\ref{s0pi}) and (\ref{d0pi}) have $S$-wave final
states; each of their widths should be a product of the final
momentum by the square of some matrix element. We can assume that
those matrix elements differ only by the suppression factor
$\epsilon$ of Eq.(\ref{supr}). Then, accounting for this suppression
and for the increased final momentum, 378~MeV/c in the decay
(\ref{d0pi}) {\it vs.} 320~MeV/c in the decay (\ref{s0pi}), we
estimate, at last,
\be \label{gs0pib} \Gamma(D_{s0}^{*\pm}\to
D_s^{\pm}\,\pi^0)= (129\pm43)~{\rm keV}\,.\ee

On the other hand, according to FOCUS~\cite{foc}, the enhancements
near the mass 2300~MeV are consistent with kinematical effects of
feed-downs from different resonant states, while the true scalar
states $D_{0}^{*0}$ and $D_{0}^{*\pm}$ have~\cite{foc}
\be M_0^*=(2407\pm41)~{\rm MeV}\,,~~~~\Gamma_0^*= (240\pm81)~{\rm
MeV}\,; \label{d0foc} \ee
\be M_{\pm}^*=(2403\pm38)~{\rm MeV}\,,~~~~\Gamma_{\pm}^*=
(283\pm42)~{\rm MeV}\,. \label{d+foc} \ee
At such parameters the decay (\ref{d0pi}) has the final momentum
459~MeV and leads to the estimate
\be \label{gs0pif} \Gamma(D_{s0}^{*\pm}
\to D_s^{\pm}\,\pi^0)= (109\pm16)~{\rm keV}\,.\ee
It is somewhat lower but consistent with the
Belle-based estimate (\ref{gs0pib}).

Let us compare now the pion decay (\ref{s0pi}) with radiative
decays of $D_{sJ}^{\pm}(2317)\,$. None of such radiative decays
has been observed yet. If the $D_{sJ}^{\pm}(2317)$
indeed has $J^P=0^+\,$, then its decay to $D_s\,$, with $J^P=0^-\,$,
is strictly forbidden (as any radiative 0-0 transition). Possible
is the decay \be D_{s0}^{*\pm}(2317)\to D_s^{*\pm}\,\gamma\,.
\label{s0rad} \ee Its absence in experiment is one of strange
properties of $D_{sJ}(2317)\,$. This decay should exist, if the
interpretation of $D_{sJ}(2317)$ as $D_{s0}^{*}$ is correct. To
estimate its expected probability, we compare it to the similar decay
\be \chi_{c0}\to J/\psi\,\gamma\,. \label{chi0} \ee

The both decays should correspond to E1 transitions. Their partial
widths may be written as \be \Gamma(0^+\to1^-+\gamma)=|\langle
{\textbf d}\rangle|^2\,p_{\gamma}^3\,. \label{rad}\ee Here
$\langle {\textbf d}\rangle$ may be interpreted as the transition
matrix element of the operator of electric dipole moment,
$p_{\gamma}$ is the momentum of the produced photon. In its turn,
the dipole moment element may be written as \be \langle {\textbf
d}\rangle=\langle\Delta e\rangle\cdot\langle{\textbf r}\rangle\,,
\label{dipm}\ee where $\langle\Delta e\rangle$ is the weighted
charge difference of constituents in the 2-body $(q\ovl q)$
system. For the charmonium $(c\ovl c)$ it is
$$\langle\Delta e\rangle_{c\ovl c}=(e_c-e_{\ovl c})/2=2/3\,.$$
For the $D_s$-system $(c\ovl s)\,$, with unequal quark masses,
we take non-relativistic expression $$\langle\Delta
e\rangle_{c\ovl s}=(m_c e_{\ovl s}-m_s
e_c)/(m_c+m_s)\,.$$ Using the naive estimation $m_c\approx3m_s\,$,
we obtain $\langle\Delta e\rangle_{c\ovl
s}\approx\langle\Delta e\rangle_{c\ovl c}/8\,$. Fitting
experimental meson spectra to some detailed potential calculations
gives $m_c\approx4m_s$ (see, \textit{e.g.}, ref.\cite{God}); then
$\langle\Delta e\rangle_{c\ovl s}\approx\langle\Delta
e\rangle_{c\ovl c}/5\,$. So we can safely take \be
\langle\Delta e\rangle_{c\ovl s}\leq\frac{1}{5}\,
\langle\Delta e\rangle_{c\ovl c}\,. \label{e-av} \ee

The matrix element $\langle{\textbf r}\rangle$ could also be
different for the systems $c\ovl c$ and $c\ovl s$ (in
the pure Coulomb case the difference would be determined by the
factor $(m_c/m_s+1)/2\,$, which can be $\sim2$ for realistic constituent
masses). However, even ground state of charmonium is not sufficiently
compact to be concentrated in the pure Coulomb area. As a result,
this matrix element should change much smaller. For simplicity, as in
other cases, we take it independent of the constituent masses, and
apply relation (\ref{e-av}) directly to matrix elements of the dipole
moment.

The partial width of decay (\ref{chi0}) is known experimentally to
be~\cite{PDG} \be \Gamma(\chi_{c0}\to J/\psi\,\gamma)
=119\pm19~{\rm keV}\,.\label{wchi0} \ee Its final photon momentum
is 303~Mev/c, {\it vs.} about 200~MeV/c  in decay (\ref{s0rad}).
Accounting for suppression (\ref{e-av}) and for the $p^3$
behaviour of Eq.(\ref{rad}) with the smaller final momentum,
we can now estimate expected partial width of the decay (\ref{s0rad})
\be \Gamma(D_{s0}^*\to D_s^*\,\gamma)\leq1.4~{\rm
keV}\,.\label{ws0g} \ee Treatment of Eq.(\ref{e-av}) as equality with
the ``naive'' coefficient 1/8 instead of ``safe'' 1/5 would give the
value \be \Gamma(D_{s0}^*\to D_s^*\,\gamma)\approx0.55~{\rm
keV}\,.\label{nws0g} \ee

 Thus we expect the ratio $\Gamma(D_{s0}^*\to
D_s^*\,\gamma)/\Gamma(D_{s0}^*\to D_s\,\pi^0)$ to be not
higher than $\sim1.5\%\,$. The present experimental boundary of
CLEO~\cite{FPort} \be \frac{{\rm Br}(D_{sJ}^*(2317)\to
D_s^*\,\gamma)}{{\rm Br}(D_{sJ}^{*}(2317)\to
D_s\,\pi^0)}<6\%\label{exp} \ee completely agrees with our
expectation.

\section{Decays of $D_{s1}(2460)$}

Here we begin with considering the radiative decay \be
D_{s1}(2460)\to D_s\,\gamma\,,\label{s1rad}\ee which was
definitely observed by Belle~\cite{Bel2}. The $D_s$-meson has
$J^P=0^-$, and for $D_{s1}(2460)$ we assume $J^P=1^+$. Then
the radiative decay (\ref{s1rad}) corresponds to E1 transition,
and can be estimated in essentially the same way as done in the
preceding section for the radiative decay of $D_{s0}^*\,$.

Charmonium analog of $D_s$ is $\eta_c$ with $J^{PC}=0^{-+}$.
The state $D_{s1}(2460)$ has two possible analogs, $h_c$ with
$J^{PC}=1^{+-}$, and $\chi_{c1}$ with $J^{PC}=1^{++}$. Decay
$\chi_{c1}\to\eta_c\,\gamma$ is forbidden by $C$-parity, and thus,
analogous to decay (\ref{s1rad}) could be only the radiative decay
$h_c\to\eta_c\,\gamma\,$.  However, the state $h_c$ itself is badly
known (see Listings in tables~\cite{PDG}), and its radiative decay
has never been observed. Instead, we again can use decay (\ref{chi0}),
but with an essential note. Contrary to decays (\ref{s0rad}) and
(\ref{chi0}), which are $0^+\to1^-$ transitions, decay (\ref{s1rad})
corresponds to the transition $1^+\to0^-$.  This leads to the
radiative width \be \Gamma(1^+\to0^-+\gamma)=\frac13\,|\langle
{\textbf d}\rangle|^2\,p_{\gamma}^3\,, \ee which should be compared
with the expression (\ref{rad}). The new factor 1/3 appears due to
averaging over spin 1 states of the initial system, instead of
summing over them for the final system in expression (\ref{rad}).
Taking the other factors to be the same as before, and accounting
for the larger final momentum in decay (\ref{s1rad}), about 440~MeV/c,
we obtain the upper boundary \be \Gamma(D_{s1}\to D_s\,\gamma)
\leq5~{\rm keV}\,.\label{ws1g} \ee It would give \be \Gamma(D_{s1}
\to D_s\,\gamma)\approx2~{\rm keV} \label{nws1g}\ee at the ``naive''
treatment of Eq.(\ref{e-av}) as an equality with the coefficient 1/8,
instead of 1/5.

Pionic decay
\be D_{s1}(2460)\to D_s^*\,\pi^0\,,\label{s1pi}\ee
with $p_{\pi}=425$~MeV/c, violates isospin symmetry. Its
theoretical consideration may appear ambiguous since there are
two possible pion transitions $1^+\to1^-\,$, through $S$- or
$D$-wave. Of course, they can be discriminated experimentally,
but this has not been done yet for $D_{s1}(2460)\,$.

In the charmed-nonstrange sector there are two
$1^+$-states~\cite{Bel3,foc,PDG} with nearby masses. One of them
is relatively narrow, another is much wider. Those are $D_1$
with \be M_{D_1}=(2421.4\pm1.7)~{\rm
MeV}\,,~~~\Gamma_{D_1}=(23.7\pm4.8)~{\rm MeV} \label{D1} \ee and
$D'_1$ with \be M_{D'_1}=(2427\pm36)~{\rm
MeV}\,,~~~\Gamma_{D'_1}=(384\pm120)~{\rm MeV}\,. \label{D'1} \ee
We use here the two masses and widths from the simultaneous coherent
analysis of the $D^*\pi$-system~\cite{Bel3}. PDG-tables~\cite{PDG}
contain only a narrower state, its parameters being consistent with
values (\ref{D1}).  On the other hand, analysis of the FOCUS
Collaboration~\cite{foc} reveals only a wider state, which
mass and width are consistent with (\ref{D'1}).

Further, according to Ref.\cite{Bel3}, $D_1$ decays mainly through
$D$-wave, and $D'_1$ through $S$-wave (the latter agrees with
data~\cite{foc}), thus explaining the large difference of the widths.
The mutual admixtures are about 10\% in the wave functions~\cite{Bel3},
and as the first step we neglect the mixing.

Here we need to make an assumption about partial-wave properties of
the decay (\ref{s1pi}).  If it goes (mainly) in $S$-wave, we should
compare it to the $S$-wave decay $D'_1\to D^*\,\pi^0\,,$ with the
final momentum 482~MeV/c. This can be done in exactly the same way
as for decays (\ref{s0pi}) and (\ref{d0pi}), and leads to \be
\Gamma^{(S)}(D_{s1}(2460)\to D_s^*\,\pi^0)=(187\pm73)~{\rm keV}\,.
\label{gSds1} \ee But if the decay (\ref{s1pi}) is $D$-wave one, it
should be compared to the $D$-wave decay $D_1\to D^*\,\pi^0\,$, having
the final momentum 477~MeV/c. The width for such decays may be written
as product of the 5-th power of the final momentum by the square of
a matrix element.  As before, we relate the matrix elements for
isospin-violating ($D_{s1})$ and isospin-conserving ($D_1$) decays
by the suppression parameter $\epsilon$ and then account for the
$p^5$-behaviour of the partial width. Thus we obtain \be
\Gamma^{(D)}(D_{s1}(2460)\to D_s^*\,\pi^0)=(7.4\pm2.3)~{\rm keV}\,.
\label{gDds1} \ee

To discriminate between these two possibilities, we can use the
experimental fact of observing decay (\ref{s1rad}) with relative
intensity~\cite{Bel1,Bel2}
\be \label{g/pi} \frac{\Gamma(D_{s1}(2460)\to D_s\,\gamma)}
{\Gamma(D_{s1}(2460)\to D_s^*\,\pi^0)} \sim 0.5\,. \ee
Comparison of values (\ref{ws1g}), (\ref{gSds1}) and (\ref{gDds1})
shows that our estimate for the $S$-wave case leads to strong
disagreement with the ratio (\ref{g/pi}), while it is quite
compatible with the $D$-wave case.

\section{Discussion and conclusion}

Let us summarize our assumptions and their implications.

1) First of all, we have assumed that overlap integrals in decay
amplitudes are nearly independent of the constituent quark masses.
This assumption would be incorrect if the binding potential were
Coulombic. But realistic potential should be more complicated,
thus making our assumption admissible. In any case, it is inspired
by the phenomenology of measured meson mass differences.

2) Another assumption may be less clearly motivated. It is
the universal character of the suppression factor $\epsilon$,
as given by Eq.(\ref{supr}), for
isospin-violating amplitudes in decays of mesons containing at least
one heavy quark. We will discuss this point below in more detail.

All our estimates begin with decays of mesons which are usually
considered as quark-antiquark systems (charmonium, excited $D$-mesons).
Therefore, we never use, at least explicitly, presence of 4-quark
components.

With such assumptions we are able to describe the present data on
relation of photon and pion decays of the new $D_s$ mesons, but only
if the decay (\ref{s1pi}), {\it i.e.} $D_{s1}(2460)\to D_s^*\,\pi^0$,
goes (mainly) through $D$-wave. In terms of the effective heavy quark
description, this would mean that $D_{s1}(2460)$ corresponds to the
state with $j=3/2\,$.  Though the question can (and will) be solved by
direct measurement of the angular distribution in the decay, today such
prescription disagrees with familiar expectation of the theoretical
community.  Moreover, if it is correct, another assumed $1^+$ state,
$D_{sJ}(2535)\,$, should have $j=1/2$ and decay to $D^*K$ through
$S$-wave. In that case one could hardly understand its small width
$\Gamma_{tot}<2.3$~MeV~\cite{PDG}.

If $D_{s1}(2460)$ corresponds to $j=1/2\,$, our calculations can
still be made consistent with data, if we weaken the above
assumptions. As the easiest way to this goal we can drop out
the universality of the suppression factor (\ref{supr}) for
isospin violation. Indeed, if we accept estimates (\ref{nws0g})
and (\ref{nws1g}) for radiative decays, then the amplitude for
the $S$-wave decay (\ref{s1pi}) should be suppressed by the
factor $\sim\epsilon/7\,$, instead of $\sim\epsilon\,$, to satisfy
experimental relation (\ref{g/pi}). Such suppression ($\sim6\cdot
10^{-3}$) has rather familiar order of smallness and does not look
too severe.

On the other side, to satisfy the boundary (\ref{exp}), the
amplitude for decay (\ref{s0pi}),  {\it i.e.} $D_{s0}^{*\pm}(2317)
\to D_s^{\pm}\,\pi^0$, should not be suppressed stronger than
$\sim\epsilon/4\,$, which is $\sim10^{-2}$. Thus, the isospin
violation in decays (\ref{s0pi}) and (\ref{s1pi}) seems to be
non-universal, and different from violation effects in decays of
charmonium.

This might look strange, since all the cases should have universal
contributions due to $(\eta\,\pi^0)$-mixing. Note, however, that
for charmonium we compare decays to $\eta$ and to $\pi^0$, while in
estimating decays (\ref{s0pi}) and (\ref{s1pi}) we compare pionic
decays of different systems, $c\ovl s$ and $c\ovl d\,$. Such ratios
should not be necessarily the same.

Moreover, as shown in Ref.\cite{yaa}, the universal mixing violation
of isospin  symmetry should be always accompanied by non-universal
direct violation (compare the mixing {\it vs.} direct violation of
$CP$-parity, say, in kaon decays). Due to the small mass difference,
the $(\omega\,\rho^0)$-mixing strongly enhances violation, and makes
the whole effect to be nearly universal. The enhancement due to
$(\eta\,\pi^0)$-mixing is weaker, and non-universal contributions of
direct violation may become more essential. Thus, non-universality of
isospin violation in decays of different $D_s$-mesons is not amusing.

One may also try to weaken the first of our above assumptions. It is
easy to see that some increase of the overlap integrals for
$D_s$-mesons as compared with charmonium allows to reach better
agreement between our estimates and experimental data. It could be
done even with universal suppression factor of isospin violation,
but then the increase should be faster than for the Coulomb potential,
in contradiction with data on mass differences. Therefore, the
suppression, most probably, should be non-universal. We will not
discuss here structure of the admissible effective potential.

These two modifications of our assumptions provide interesting
features. Estimates (\ref{nws0g}) and (\ref{nws1g}), appended
by non-universal isospin violation, provide for the new $D_s$
mesons the total widths of less than $\sim10$~keV. Universal
isospin-violation suppression, together with increasing overlap
integrals, would give the total widths $\sim100$~keV. So, the two
possibilities could be discriminated by special precise experiments.

As a conclusion, we can say that present data on decay properties of
new $D_s$ mesons may be described in terms of a quark-antiquark bound
system with some non-Coulombic potential. Presence of an essential
4-quark component is not required. Similar conclusion was derived
earlier~\cite{CJ} on the basis of spectroscopy only. Our additional,
and new, result is probable non-universality of isospin violation
in pionic decays of different $D_s$ states, even if it is related
to $(\eta\,\pi^0)$-mixing. Important and clarifying new information
could come from measurements of total widths for the new mesons,
though this seems to be a very hard problem.

\section*{Acknowledgments}

The work was partly supported by the Russian State Grant
SS-1124.2003.2, and by the Sofja Kovalevskaja Programme of the
Alexander von Humboldt Foundation, the German Federal Ministry of
Education and Research, and the Programme for Investment in the
Future of the German Government.

\end{document}